\documentclass[superscriptaddress,aps,preprintnumbers,amsmath,showpacs,amssymb,prd,nofootinbib,reprint]{revtex4-1}
\usepackage{amsfonts}
\usepackage[mathscr]{eucal}
\usepackage{ascmac}
\usepackage{dcolumn}
\usepackage{bm}
\usepackage[colorlinks=true,linkcolor=blue,citecolor=blue]{hyperref}
\usepackage{hyperref}
\usepackage{color}
\usepackage{amsmath}
\usepackage{graphics}
\usepackage[all,cmtip]{xy}
\usepackage{float}

\begin{document}


\def\a{\alpha}
\def\b{\beta}
\def\c{\varepsilon}
\def\d{\delta}
\def\f{\phi}
\def\g{\gamma}
\def\h{\theta}
\def\k{\kappa}
\def\l{\lambda}
\def\m{\mu}
\def\n{\nu}
\def\p{\psi}
\def\q{\partial}
\def\r{\rho}
\def\s{\varphi}
\def\t{\tau}
\def\u{\upsilon}
\def\v{\varphi}
\def\w{\omega}
\def\x{\xi}
\def\y{\eta}
\def\z{\zeta}
\def\D{\Delta}
\def\G{\Gamma}
\def\H{\Theta}
\def\L{\zeta}
\def\F{\Phi}
\def\P{\Psi}
\def\S{\Sigma}

\def\aa{{\dot \a}}
\def\bb{{\dot \b}}
\def\ss{{\bar \s}}
\def\hh{{\bar \h}}
\def\CA{{\cal A}}
\def\CB{{\cal B}}
\def\CC{{\cal C}}
\def\CD{{\cal D}}
\def\CE{{\cal E}}
\def\CG{{\cal G}}
\def\CH{{\cal H}}
\def\CI{{\cal I}}
\def\CK{{\cal K}}
\def\CL{{\cal L}}
\def\CR{{\cal R}}
\def\CM{{\cal M}}
\def\CN{{\cal N}}
\def\CO{{\cal O}}
\def\CP{{\cal P}}
\def\CQ{{\cal Q}}
\def\CS{{\cal S}}
\def\CT{{\cal T}}
\def\CW{{\cal W}}

\newcommand{\Slash}[1]{{\ooalign{\hfil/\hfil\crcr$#1$}}}

\def\o{\over}
\newcommand{\gsim}{ \mathop{}_{\textstyle \sim}^{\textstyle >} }
\newcommand{\lsim}{ \mathop{}_{\textstyle \sim}^{\textstyle <} }
\newcommand{\vev}[1]{ \left\langle {#1} \right\rangle }
\newcommand{\bra}[1]{ \langle {#1} | }
\newcommand{\ket}[1]{ | {#1} \rangle }
\newcommand{\EV}{ {\rm eV} }
\newcommand{\KEV}{ {\rm keV} }
\newcommand{\MEV}{ {\rm MeV} }
\newcommand{\GEV}{ {\rm GeV} }
\newcommand{\TEV}{ {\rm TeV} }
\def\diag{\mathop{\rm diag}\nolimits}
\def\Spin{\mathop{\rm Spin}}
\def\SO{\mathop{\rm SO}}
\def\O{\mathop{\rm O}}
\def\SU{\mathop{\rm SU}}
\def\U{\mathrm{U}}
\def\Sp{\mathop{\rm Sp}}
\def\USp{\mathop{\rm USp}}
\def\SL{\mathop{\rm SL}}
\def\tr{\mathop{\rm tr}}
\def\rank{\mathop{\rm rank}}

\def\spin{\mathfrak{ spin}}
\def\so{\mathfrak{so}}
\def\o{\mathfrak{o}}
\def\su{\mathfrak{su}}
\def\u{\mathfrak{u}}
\def\sp{\mathfrak{sp}}
\def\usp{\mathfrak{usp}}
\def\sl{\mathfrak{sl}}
\def\e{\mathfrak{e}}

\def\beq#1\eeq{\begin{align}#1\end{align}}
\def\alert#1{{\color{red}[#1]}}


\preprint{
}

\title{
New  $\mathcal{N}=2$ dualities
}

\author{
Dan Xie
}
\affiliation{Center of Mathematical Sciences and Applications, Harvard University, Cambridge, 02138, USA}
\affiliation{Jefferson Physical Laboratory, Harvard University, Cambridge, MA 02138, USA}

\author{
Shing-Tung Yau
}
\affiliation{Department of Mathematics, Harvard University, Cambridge, MA 02138, USA}


\begin{abstract} 
We consider $\mathcal{N}=2$ superconformal field theory with following properties: a) Coulomb branch operators have fractional scaling dimensions, b) there are exact marginal deformations . The weakly coupled gauge theory descriptions are found by decomposing 3d mirror into different components, and different decompositions correspond to different duality frames. The gauge theory is formed by gauging Argyres-Douglas matter, and
we write down all duality frames for several classes with infinite sequence of theories. 
\end{abstract}

\maketitle


\section{Introduction
\label{sec:introduction}}
Dualities of superconformal field theories have played a central role in our understanding of their dynamics.
The typical examples are S dualities of four dimensional $\mathcal{N}=4$ Super Yang-Mills theory \cite{Montonen:1977sn} and $\mathcal{N}=2$ 
$SU(2)$ gauge theory with $N_f=4$ fundamental flavors \cite{Seiberg:1994aj}. Argyres-Seiberg \cite{Argyres:2007cn} found a remarkable generalization of these dualities by 
showing that the dual of  $\mathcal{N}=2$ $SU(3)$ gauge theory with $N_f=6$ is $SU(2)$ theory coupled to $E_6$ matter and a fundamental hypermultiplet, 
later Gaiotto found a vast generalization of such dualities by using M5 brane constructions \cite{Gaiotto:2009we}. 

The special property of the above theories is that the Coulomb branch operators have integral scaling dimensions. 
There is a much larger class of $\mathcal{N}=2$ SCFTs which have fractional scaling dimensions and exact marginal deformations \cite{Argyres:1995jj,Xie:2012hs,Wang:2015mra,Xie:2015rpa} (we call them Argyres-Douglas (AD) theories), and 
one may wonder whether they also have interesting S dualities. Indeed, two very interesting examples 
have been found in \cite{Buican:2014hfa}, and later a class of self-dual theories are found using compactification of 
6d $(1,0)$ theories in \cite{DelZotto:2015rca}. However, the S duality properties of most of these theories have not been studied. 

The purpose of this letter is to find S duality for above mentioned theories. We first revisit S duality of theories considered by Gaiotto \cite{Gaiotto:2009we}, 
and reinterpret it in terms of decompositions of the 3d mirror. Since many Argyres-Douglas theories do have 3d mirrors \cite{Xie:2012hs}, we may wonder 
if S duality also corresponds to different decompositions of the 3d mirror, and it is found that such procedure works well and 
in a much more non-trivial way. We apply this method to many classes of infinite sequence of theories and find all the duality frames. 
These new dualities exhibit many new features, which we hope can help us better understand  dynamics of quantum field theories.

\section{Review of Gaiotto duality }
A large class of $\mathcal{N}=2$ SCFTs can be engineered by putting 6d $(2,0)$ theory on a Riemann surface with regular punctures. 
The S duality of these theories are interpreted as different decompositions of punctured Riemann surface into three punctured spheres \cite{Gaiotto:2009we}. 

We would like to interpret S duality of these theories in terms of different decompositions of its 3d mirror. 
The 3d mirror of above theories is a star shaped quiver which has a single central node and legs (quiver tail) given by punctures \cite{Benini:2010uu}. 
The Higgs branch of 4d theory is equivalent to Coulomb branch of its 3d mirror. The following two facts about 3d quiver we have are important to us: a)
 A node is called \textbf{balanced} if $n_f=2n_c$, and a quiver is called \textbf{good} if $n_f\geq 2n_c$ for all  quiver nodes \cite{Gaiotto:2008ak}; b) if there is a \textbf{balanced} subquiver with ADE shape, then
 there is a corresponding ADE flavor group on the Coulomb branch. We always assume the quiver is good as there is a way to turn a bad quiver into a good quiver \cite{Nanopoulos:2010bv}.

\begin{figure}[H]
\centering
  \includegraphics{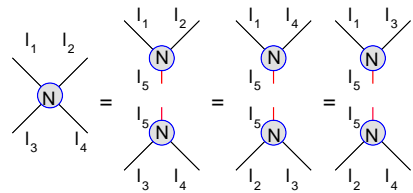}
  \caption{S duality of a SCFT defined by fourth punctured sphere is interpreted as different decompositions of its  mirror.}
  \label{gaiotto}
\end{figure}

Let's focus on theory defined by a fourth punctured sphere. 
We could reinterpret the S duality as three ways of decomposing the mirror into two 
star-shaped quivers with 3 legs (see figure.\ref{gaiotto}), and the gauging process corresponds to eliminating new legs carrying non-abelian symmetries on the Coulomb branch and merging 
central nodes. The central task of finding S duality is to find the new legs appearing in  decompositions. 

\section{Argyres-Douglas matter}
Argyres-Douglas theories can be engineered by putting 6d $(2,0)$ theory on a Riemann sphere with irregular singularity \cite{Gaiotto:2009hg,Xie:2012hs,Wang:2015mra}. To define 
a SCFT, one can have the following two types of configurations: a) a single irregular singularity; b) an irregular singularity and 
a regular singularity. For 6d $A_{N-1}$ theory, the irregular singularities have been classified in \cite{Xie:2012hs} using the classification of irregular singularity of 
Hitchin's equation:
\begin{align}
& \text{I}_{k,N}:~~~\Phi={T\over z^{{k\over N}+2}}+\ldots,~~ \text{II}_{k,N}:~~\Phi={T\over z^{{k\over N-1}+2}}+\ldots  \nonumber\\
&\text{III}_{Y_n,\ldots, Y_1}:~~\Phi={T_n\over z^n}+\ldots+{T_1\over z},~Y_n\subseteq Y_{n-1}\subseteq \dots \subseteq Y_1
\end{align}
Here $Y_i$ denotes the Young Tableaux describing degeneracy of the eigenvalues of $T_i$. The AD theory defined 
by them are called type I, II and III theories\footnote {Type I theory is first discovered in \cite{Cecotti:2010fi} and called $(A_{k-1},A_{N-1})$ theory.}. 
Type IV theory is defined by an irregular singularity and a regular singularity, and it is denoted as $(P, Y)$\footnote{The integer $k$  take values $k>-N$ for $(\text{I},Y)$ case, and $k>-N+1$ for $(\text{II},Y)$ case . When $Y$ is a full puncture $F$, the theory 
$(I_{k,N},F)$ is also called $D_{N+k}(SU(N)$ theory in \cite{Cecotti:2012jx}.}, where $P$ specifies irregular singularity and  $Y$ 
specifies  regular singularity \cite{Gaiotto:2009we}. The Seiberg-Witten (SW) curve and Coulomb branch spectrum for these theories are described in \cite{Xie:2012hs}, i.e. the SW curve is identified with the
spectral curve of corresponding Hitchin's system:
\begin{equation}
x^N+\sum_{i=2}^N\phi_i(z)x^{N-i}=0.
\end{equation}
The major new features about these theories are: a): the scaling dimensions of Coulomb branch
operators are fractional; b): there are other dimensional coupling constants besides mass deformations. 

The type III and type IV theory are particularly interesting for this paper: a) they can admit non-abelian flavor symmetry which
can be gauged to form  superconformal gauge theory, b): they can have exact marginal deformations.    

We define Argyres-Douglas matter as those \textbf{isolated} SCFTs with the following properties: a): the scaling dimension 
of Coulomb branch operators are \textbf{fractional}; b): the theory carries \textbf{non-abelian} flavor symmetry.
 The non-abelian flavor symmetry of type IV theory is carried by the regular singularity, and that 
of type III theory can be found using its 3d mirror. 

\subsection{flavor central charge}
The flavor symmetry of an Argyres-Douglas matter has the form $G_1\times G_2\ldots \times G_n$, and
we need to know the flavor central charge $k$ for various non-abelian factors. These numbers are computed for 
type IV theory if $Y$ is taken to be the full puncture \cite{Xie:2013jc}, and the answer is 
\begin{align}
&(\text{I}_{k,N}, F):~~k_F={N(N+k-1)\over N+k},~~\nonumber\\
&(\text{II}_{k,N}, F):~~k_F={kN+(N-1)^2\over N+k-1}.~~~~
\end{align}
They are found by identifying $k_G$ as the maximal scaling dimension of the Coulomb branch operators.
We would like to generalize the computation of $k_G$ to arbitrary regular puncture. 

\begin{figure}[H]
\centering
  \includegraphics{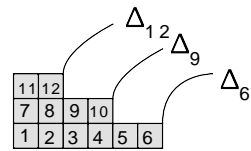}
  \caption{Here $Y=[3,3,2,2,1,1]$, and the boxes of Young Tableaux are labeled. The edge boxes are also indicated. }
  \label{young}
\end{figure}

Let's consider a regular puncture with Young Tableaux $(r_1^{n_1},r_2^{n_2},\ldots, r_s^{n_s})$ with $r_1>..>r_s$, then the flavor symmetry 
is $G=(\cup_{i=1}^s \text{U}(n_i))/\text{U}(1)$ with non-abelian factor $SU(n_1)\times \ldots\times SU(n_i)$. 
To determine the flavor central charge, let's first draw the Young Tableaux such that the columns are $r_1,..,r_1,r_2,..,r_2,\ldots$,
 then label the boxes of $Y$ from 1 to N row by row starting from bottom left.  
The order of pole of $i$th differential $\phi_i$ at the regular puncture is $i-s_i$ with $s_i$ 
the height of $i$th box. The boxes at the right edges
are special, and we record their labels as $(a_1,\ldots, a_r)$. We record the scaling dimension of the operator associated with monomials ${O_{a_i}\over z^{a_i-s_{a_i}}}x^{N-i}$:
$(\Delta_{a_1},\ldots, \Delta_{a_r})$, see figure. \ref{young}. Let's now consider a non-abelian factor $SU(n_i)$(associated with columns of $Y$ with height $r_i$) and assume that the 
maximal label for those columns are $A_i$, then we have  
\begin{equation}
k_{SU(n_i)}=\sum_{a_j\leq A_i}\Delta_{a_j}.
\end{equation}
For the full puncture $Y=[1,\ldots,1]$, we have $A=N$  and there is only one edge box with label $N$. So $k_G=\Delta_{N}$ which 
is actually the maximal scaling dimension in the spectrum.
Notice that the maximal scaling dimension is a local concept associated with  regular puncture, globally, there might 
be no such Coulomb branch operators, and this resolves a puzzle raised in \cite{Buican:2014hfa}. The above procedure can also be generalized to type III theories \cite{dxie:201601} using
the isomorphism between type III theories and type IV theories.

The flavor central charge of a subgroup $G{'}$ of a non-abelian simple flavor group $G$
can be calculated using group representation theory as described in \cite{Argyres:1995jj}. One useful result is that
for a $G^{'}=SU(n)$ subgroup of a $G=SU(n+m)$ flavor group, the flavor central charges satisfy the condition
$k_{G^{'}}=k_G$.

If we gauge a flavor  group $G$ of an AD matter, the flavor central charge $k_G$ is equal to the contribution to the $\beta$ function.
One can form superconformal gauge theory by choosing AD matters appropriately. 
Given the rich choices of 
AD matter, there are many new interesting  SCFTs one can construct.  In this paper, we take a different route by focusing on  AD
theories with exact marginal deformations and try to find its gauge theory descriptions involving gauging AD matters.

\section{S duality}
Many of type III and type IV theories  have exact marginal deformations which are 
associated with the parameters in the leading order matrix $T_n$ specifying irregular singularity. 
We would like to find out the weakly coupled gauge theory descriptions for them. 
These theories do admit 3d mirror, and one might wonder if we can find different duality frames 
by decomposing the 3d mirror into different components, just as what we did for Gaiotto duality. In this section, we will show that 
this procedure indeed works well.  

Let's first review how to find 3d mirror for type III theories \cite{boalch2008irregular,Xie:2012hs}: Step I: Start with  Young Tableaux $Y_n=[n_1,\ldots,n_s]$,
and draw $s$ quiver nodes with ranks $n_1,\ldots,n_s$, then draw $n-2$ edges between each pair; Step II: Look at $Y_{n-1}$, 
if a column $n_i$ of $Y_n$ is  split as $[n_{1i},n_{2i},\ldots, n_{ri}]$, we split the initial quiver node $n_i$ into quiver nodes with 
rank $n_{1i},\ldots,n_{ri}$, and draw $n-3$ edges between each pair; Step III: continuing step two until $Y_1$, for $Y_1$ we do not split the quiver nodes of $Y_2$ any more, but
add a quiver tail to the split column of $Y_2$. 
The initial quiver nodes $n_1,\ldots, n_s$ (each node might split latter, but we group those quiver nodes which have the same origin in $Y_1$ as a single set) are called \textbf{core} of this quiver. 
For theory $(\text{III}, Y)$, we just attach a quiver tail of Y to the core of irregular singularity. 

The 3d mirror has many uses: a): one can read the corresponding M5 configurations and there are usually 
more than one realizations of the single 3d mirror; b): One can find out flavor symmetries of the original 4d SCFT. 

\begin{figure}[H]
\centering
  \includegraphics{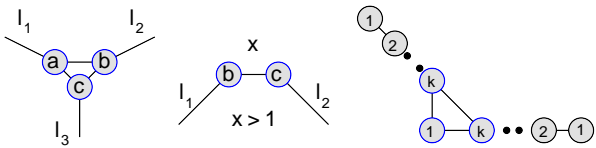}
  \caption{Left: 3d mirror for one type of AD matter; Middle: 3d mirror for other AD matters. Right: The 3d mirror for $D_2(SU(2k+1))$ theory, with $k_{SU(2k+1)}=k+{1\over 2}$. }
  \label{d2}
\end{figure}

The AD matter typically has 
a three node core if the maximal quiver edges are one, and has a two node core if the maximal quiver edges is bigger than one, 
see figure.\ref{d2}. For the cases shown in figure.\ref{d2},  If the quiver tail attached to core node with rank $a$ has maximal flavor symmetry, the flavor central charge is $k_G=a+{1\over 2}$ for the left AD matter, 
and $k_G=a+{1\over x+1}$ for the middle AD matter. 

\begin{figure}[H]
\centering
  \includegraphics{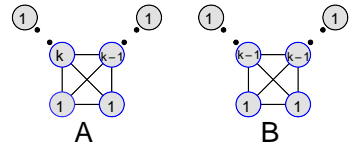}
  \caption{3d mirror for class A and class B theories. }
  \label{p1}
\end{figure}

Let's consider  following two classes of type III theories whose defining data are
listed in table.\ref{t1}. The 3d mirror for these theories are depicted in figure.\ref{p1}. 

\begin{table}[t]
\centering
\begin{tabular}{|c|c|c|c|}
\hline
& $Y_3$ & $Y_2$ & $Y_1$ \\ \hline
$A$ & $[k,k-1,1,1]$ & $[k,k-1,1,1]$ & $[1,\ldots,1]$  \\ \hline
$B$ & $[k-1,k-1,1,1]$ & $[k-1,k-1,1,1]$ & $[1,\ldots,1]$ \\ \hline
\end{tabular}
\caption{Defining data for class A and class B theories.}
\label{t1}
\end{table}

\begin{figure}[H]
\centering
  \includegraphics{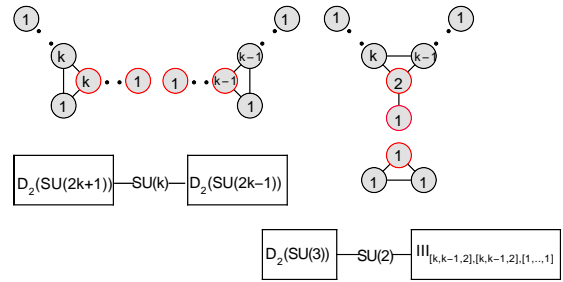}
  \caption{Decompositions of 3d mirror of class A theory and the corresponding gauge theory. }
  \label{p2}
\end{figure}

\begin{figure}[H]
\centering
  \includegraphics{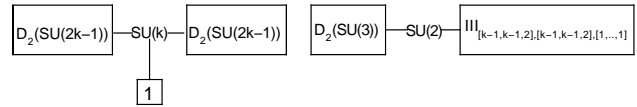}
  \caption{Different duality frames of class B theory. }
  \label{p3}
\end{figure}

Let's now study the decompositions of its 3d mirror. Let's focus 
on class A theories, and the core of its 3d mirror has rank $(k, k-1, 1,1)$ as indicated in figure.\ref{p1}.
The decompositions are regarded as decomposing the core into two sets. There are 
three such decompositions (two of them are isomorphic), see figure.\ref{p2}. 
We can write down the corresponding gauge theory descriptions by recognizing that
the decomposed components are the 3d mirror for AD matter shown in figure.\ref{p2}.
One can match various physical quantities such as the Coulomb branch spectrum of two gauge theory descriptions with the original 
SCFT.
This procedure can be easily generalized to class B theories, and the two gauge theory descriptions 
are shown in figure. \ref{p3}. We leave the verification to interested reader. 

\section{Generalizations}
Using the idea described in last section, we can find many new interesting S dualities.
Here  only some simple examples are given with full stories left for another publication.

\begin{figure}[H]
\centering
  \includegraphics{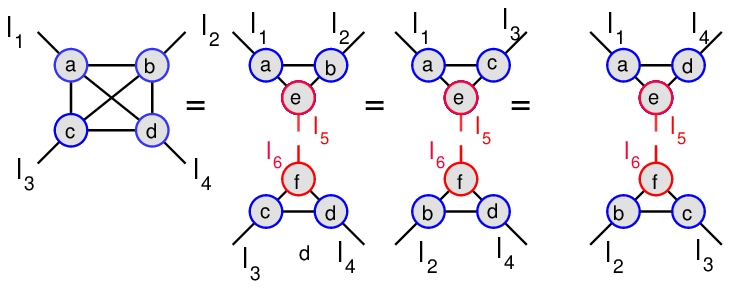}
  \caption{Three duality frames for theory whose 3d mirror has   a four node core and has only single quiver edge. }
  \label{four}
\end{figure}

\textbf{Four core quiver with single edge}:
Let's consider the SCFT whose 
3d mirror has single edge and whose core has four sets. There are
usually three decompositions, see figure. \ref{four}. The interesting question is to identify the 
new quiver tail appearing in decomposition. 

\begin{figure}[H]
\centering
  \includegraphics{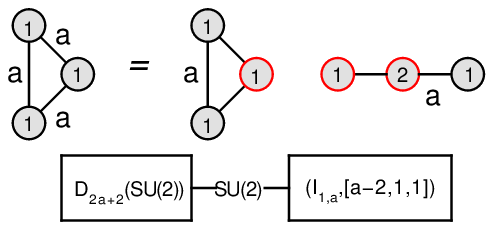}
  \caption{The decomposition of 3d mirror of $\text{I}_{3a,3}$ theory and its gauge theory description. }
  \label{multiple}
\end{figure}

\textbf{3d mirror with multiple edges}:
We now consider SCFT whose 3d mirror has quiver edges bigger than one. 
3d mirror of SCFT with one exact marginal deformation has three node core. We can 
still find the S duality by decomposing the 3d mirror. 
As an illustration, let's consider theory $\text{I}_{3a,3}$ whose 3d mirror and decompositions
are shown in figure.\ref{multiple}. There are three isomorphic 
duality frames, and the gauge theory descriptions are shown in figure.\ref{multiple} (When $a=2$, the left sub-quiver is bad, and we can reduce it to a good quiver of 
$D_{3}(SU(2))$ theory. An extra fundamental to $SU(2)$ gauge group is needed.). 

\begin{figure}[H]
\centering
  \includegraphics{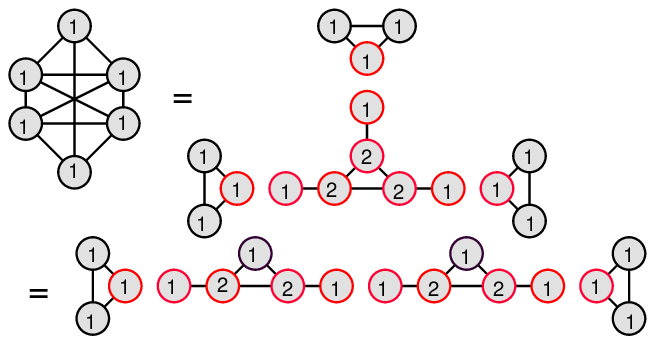}
  \caption{The two different decompositions of 3d mirror of $\text{I}_{6,6}$ theory. }
  \label{quiver1}
\end{figure}

\begin{figure}[H]
\centering
  \includegraphics{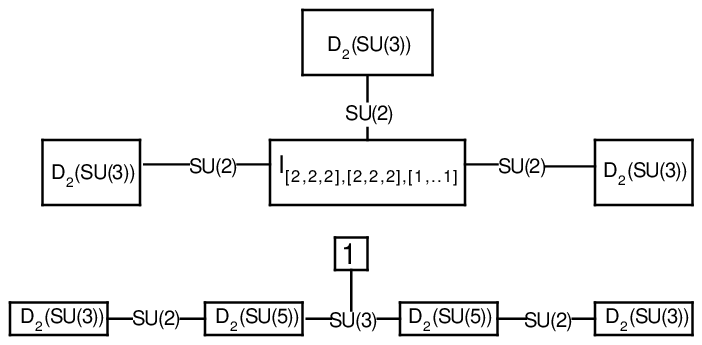}
  \caption{The two duality frames for $\text{I}_{6,6}$ theory. }
  \label{quiver2}
\end{figure}

\textbf{Generalized quiver}:
S duality of SCFT with more than one exact marginal deformations can be found using the S duality of theory with a single exact marginal deformation.
Here we only give an example. Let's consider $\text{I}_{6,6}$ theory which has three exact marginal deformations, and 
its 3d mirror is shown in figure.\ref{quiver1}.
 The two duality frames are found in figure.\ref{quiver2} by using different decompositions of its 3d mirror (see figure.\ref{quiver1}).

%
\section*{Acknowledgments}
The work of S.T Yau is supported by  NSF grant  DMS-1159412, NSF grant PHY-
0937443, and NSF grant DMS-0804454.  
The work of DX is supported by Center for Mathematical Sciences and Applications at Harvard University.
%

\vspace{1cm}

\bibliographystyle{apsrev4-1}
\bibliography{ref}

\begin{thebibliography}{19}%
\makeatletter
\providecommand \@ifxundefined [1]{%
 \@ifx{#1\undefined}
}%
\providecommand \@ifnum [1]{%
 \ifnum #1\expandafter \@firstoftwo
 \else \expandafter \@secondoftwo
 \fi
}%
\providecommand \@ifx [1]{%
 \ifx #1\expandafter \@firstoftwo
 \else \expandafter \@secondoftwo
 \fi
}%
\providecommand \natexlab [1]{#1}%
\providecommand \enquote  [1]{``#1''}%
\providecommand \bibnamefont  [1]{#1}%
\providecommand \bibfnamefont [1]{#1}%
\providecommand \citenamefont [1]{#1}%
\providecommand \href@noop [0]{\@secondoftwo}%
\providecommand \href [0]{\begingroup \@sanitize@url \@href}%
\providecommand \@href[1]{\@@startlink{#1}\@@href}%
\providecommand \@@href[1]{\endgroup#1\@@endlink}%
\providecommand \@sanitize@url [0]{\catcode `\\12\catcode `\$12\catcode
  `\&12\catcode `\#12\catcode `\^12\catcode `\_12\catcode `\%12\relax}%
\providecommand \@@startlink[1]{}%
\providecommand \@@endlink[0]{}%
\providecommand \url  [0]{\begingroup\@sanitize@url \@url }%
\providecommand \@url [1]{\endgroup\@href {#1}{\urlprefix }}%
\providecommand \urlprefix  [0]{URL }%
\providecommand \Eprint [0]{\href }%
\providecommand \doibase [0]{http://dx.doi.org/}%
\providecommand \selectlanguage [0]{\@gobble}%
\providecommand \bibinfo  [0]{\@secondoftwo}%
\providecommand \bibfield  [0]{\@secondoftwo}%
\providecommand \translation [1]{[#1]}%
\providecommand \BibitemOpen [0]{}%
\providecommand \bibitemStop [0]{}%
\providecommand \bibitemNoStop [0]{.\EOS\space}%
\providecommand \EOS [0]{\spacefactor3000\relax}%
\providecommand \BibitemShut  [1]{\csname bibitem#1\endcsname}%
\let\auto@bib@innerbib\@empty
\bibitem [{\citenamefont {Montonen}\ and\ \citenamefont
  {Olive}(1977)}]{Montonen:1977sn}%
  \BibitemOpen
  \bibfield  {author} {\bibinfo {author} {\bibfnamefont {C.}~\bibnamefont
  {Montonen}}\ and\ \bibinfo {author} {\bibfnamefont {D.~I.}\ \bibnamefont
  {Olive}},\ }\href {\doibase 10.1016/0370-2693(77)90076-4} {\bibfield
  {journal} {\bibinfo  {journal} {Phys. Lett.}\ }\textbf {\bibinfo {volume}
  {B72}},\ \bibinfo {pages} {117} (\bibinfo {year} {1977})}\BibitemShut
  {NoStop}%
\bibitem [{\citenamefont {Seiberg}\ and\ \citenamefont
  {Witten}(1994)}]{Seiberg:1994aj}%
  \BibitemOpen
  \bibfield  {author} {\bibinfo {author} {\bibfnamefont {N.}~\bibnamefont
  {Seiberg}}\ and\ \bibinfo {author} {\bibfnamefont {E.}~\bibnamefont
  {Witten}},\ }\href {\doibase 10.1016/0550-3213(94)90214-3} {\bibfield
  {journal} {\bibinfo  {journal} {Nucl. Phys.}\ }\textbf {\bibinfo {volume}
  {B431}},\ \bibinfo {pages} {484} (\bibinfo {year} {1994})},\ \Eprint
  {http://arxiv.org/abs/hep-th/9408099} {arXiv:hep-th/9408099 [hep-th]}
  \BibitemShut {NoStop}%
\bibitem [{\citenamefont {Argyres}\ and\ \citenamefont
  {Seiberg}(2007)}]{Argyres:2007cn}%
  \BibitemOpen
  \bibfield  {author} {\bibinfo {author} {\bibfnamefont {P.~C.}\ \bibnamefont
  {Argyres}}\ and\ \bibinfo {author} {\bibfnamefont {N.}~\bibnamefont
  {Seiberg}},\ }\href {\doibase 10.1088/1126-6708/2007/12/088} {\bibfield
  {journal} {\bibinfo  {journal} {JHEP}\ }\textbf {\bibinfo {volume} {12}},\
  \bibinfo {pages} {088} (\bibinfo {year} {2007})},\ \Eprint
  {http://arxiv.org/abs/0711.0054} {arXiv:0711.0054 [hep-th]} \BibitemShut
  {NoStop}%
\bibitem [{\citenamefont {Gaiotto}(2012)}]{Gaiotto:2009we}%
  \BibitemOpen
  \bibfield  {author} {\bibinfo {author} {\bibfnamefont {D.}~\bibnamefont
  {Gaiotto}},\ }\href {\doibase 10.1007/JHEP08(2012)034} {\bibfield  {journal}
  {\bibinfo  {journal} {JHEP}\ }\textbf {\bibinfo {volume} {08}},\ \bibinfo
  {pages} {034} (\bibinfo {year} {2012})},\ \Eprint
  {http://arxiv.org/abs/0904.2715} {arXiv:0904.2715 [hep-th]} \BibitemShut
  {NoStop}%
\bibitem [{\citenamefont {Argyres}\ and\ \citenamefont
  {Douglas}(1995)}]{Argyres:1995jj}%
  \BibitemOpen
  \bibfield  {author} {\bibinfo {author} {\bibfnamefont {P.~C.}\ \bibnamefont
  {Argyres}}\ and\ \bibinfo {author} {\bibfnamefont {M.~R.}\ \bibnamefont
  {Douglas}},\ }\href {\doibase 10.1016/0550-3213(95)00281-V} {\bibfield
  {journal} {\bibinfo  {journal} {Nucl. Phys.}\ }\textbf {\bibinfo {volume}
  {B448}},\ \bibinfo {pages} {93} (\bibinfo {year} {1995})},\ \Eprint
  {http://arxiv.org/abs/hep-th/9505062} {arXiv:hep-th/9505062 [hep-th]}
  \BibitemShut {NoStop}%
\bibitem [{\citenamefont {Xie}(2013)}]{Xie:2012hs}%
  \BibitemOpen
  \bibfield  {author} {\bibinfo {author} {\bibfnamefont {D.}~\bibnamefont
  {Xie}},\ }\href {\doibase 10.1007/JHEP01(2013)100} {\bibfield  {journal}
  {\bibinfo  {journal} {JHEP}\ }\textbf {\bibinfo {volume} {01}},\ \bibinfo
  {pages} {100} (\bibinfo {year} {2013})},\ \Eprint
  {http://arxiv.org/abs/1204.2270} {arXiv:1204.2270 [hep-th]} \BibitemShut
  {NoStop}%
\bibitem [{\citenamefont {Wang}\ and\ \citenamefont
  {Xie}(2015)}]{Wang:2015mra}%
  \BibitemOpen
  \bibfield  {author} {\bibinfo {author} {\bibfnamefont {Y.}~\bibnamefont
  {Wang}}\ and\ \bibinfo {author} {\bibfnamefont {D.}~\bibnamefont {Xie}},\
  }\href@noop {} {\  (\bibinfo {year} {2015})},\ \Eprint
  {http://arxiv.org/abs/1509.00847} {arXiv:1509.00847 [hep-th]} \BibitemShut
  {NoStop}%
\bibitem [{\citenamefont {Xie}\ and\ \citenamefont {Yau}(2015)}]{Xie:2015rpa}%
  \BibitemOpen
  \bibfield  {author} {\bibinfo {author} {\bibfnamefont {D.}~\bibnamefont
  {Xie}}\ and\ \bibinfo {author} {\bibfnamefont {S.-T.}\ \bibnamefont {Yau}},\
  }\href@noop {} {\  (\bibinfo {year} {2015})},\ \Eprint
  {http://arxiv.org/abs/1510.01324} {arXiv:1510.01324 [hep-th]} \BibitemShut
  {NoStop}%
\bibitem [{\citenamefont {Buican}\ \emph {et~al.}(2015)\citenamefont {Buican},
  \citenamefont {Giacomelli}, \citenamefont {Nishinaka},\ and\ \citenamefont
  {Papageorgakis}}]{Buican:2014hfa}%
  \BibitemOpen
  \bibfield  {author} {\bibinfo {author} {\bibfnamefont {M.}~\bibnamefont
  {Buican}}, \bibinfo {author} {\bibfnamefont {S.}~\bibnamefont {Giacomelli}},
  \bibinfo {author} {\bibfnamefont {T.}~\bibnamefont {Nishinaka}}, \ and\
  \bibinfo {author} {\bibfnamefont {C.}~\bibnamefont {Papageorgakis}},\ }\href
  {\doibase 10.1007/JHEP02(2015)185} {\bibfield  {journal} {\bibinfo  {journal}
  {JHEP}\ }\textbf {\bibinfo {volume} {02}},\ \bibinfo {pages} {185} (\bibinfo
  {year} {2015})},\ \Eprint {http://arxiv.org/abs/1411.6026} {arXiv:1411.6026
  [hep-th]} \BibitemShut {NoStop}%
\bibitem [{\citenamefont {Del~Zotto}\ \emph {et~al.}(2015)\citenamefont
  {Del~Zotto}, \citenamefont {Vafa},\ and\ \citenamefont
  {Xie}}]{DelZotto:2015rca}%
  \BibitemOpen
  \bibfield  {author} {\bibinfo {author} {\bibfnamefont {M.}~\bibnamefont
  {Del~Zotto}}, \bibinfo {author} {\bibfnamefont {C.}~\bibnamefont {Vafa}}, \
  and\ \bibinfo {author} {\bibfnamefont {D.}~\bibnamefont {Xie}},\ }\href
  {\doibase 10.1007/JHEP11(2015)123} {\bibfield  {journal} {\bibinfo  {journal}
  {JHEP}\ }\textbf {\bibinfo {volume} {11}},\ \bibinfo {pages} {123} (\bibinfo
  {year} {2015})},\ \Eprint {http://arxiv.org/abs/1504.08348} {arXiv:1504.08348
  [hep-th]} \BibitemShut {NoStop}%
\bibitem [{\citenamefont {Benini}\ \emph {et~al.}(2010)\citenamefont {Benini},
  \citenamefont {Tachikawa},\ and\ \citenamefont {Xie}}]{Benini:2010uu}%
  \BibitemOpen
  \bibfield  {author} {\bibinfo {author} {\bibfnamefont {F.}~\bibnamefont
  {Benini}}, \bibinfo {author} {\bibfnamefont {Y.}~\bibnamefont {Tachikawa}}, \
  and\ \bibinfo {author} {\bibfnamefont {D.}~\bibnamefont {Xie}},\ }\href
  {\doibase 10.1007/JHEP09(2010)063} {\bibfield  {journal} {\bibinfo  {journal}
  {JHEP}\ }\textbf {\bibinfo {volume} {09}},\ \bibinfo {pages} {063} (\bibinfo
  {year} {2010})},\ \Eprint {http://arxiv.org/abs/1007.0992} {arXiv:1007.0992
  [hep-th]} \BibitemShut {NoStop}%
\bibitem [{\citenamefont {Gaiotto}\ and\ \citenamefont
  {Witten}(2009)}]{Gaiotto:2008ak}%
  \BibitemOpen
  \bibfield  {author} {\bibinfo {author} {\bibfnamefont {D.}~\bibnamefont
  {Gaiotto}}\ and\ \bibinfo {author} {\bibfnamefont {E.}~\bibnamefont
  {Witten}},\ }\href {\doibase 10.4310/ATMP.2009.v13.n3.a5} {\bibfield
  {journal} {\bibinfo  {journal} {Adv. Theor. Math. Phys.}\ }\textbf {\bibinfo
  {volume} {13}},\ \bibinfo {pages} {721} (\bibinfo {year} {2009})},\ \Eprint
  {http://arxiv.org/abs/0807.3720} {arXiv:0807.3720 [hep-th]} \BibitemShut
  {NoStop}%
\bibitem [{\citenamefont {Nanopoulos}\ and\ \citenamefont
  {Xie}(2011)}]{Nanopoulos:2010bv}%
  \BibitemOpen
  \bibfield  {author} {\bibinfo {author} {\bibfnamefont {D.}~\bibnamefont
  {Nanopoulos}}\ and\ \bibinfo {author} {\bibfnamefont {D.}~\bibnamefont
  {Xie}},\ }\href {\doibase 10.1007/JHEP05(2011)071} {\bibfield  {journal}
  {\bibinfo  {journal} {JHEP}\ }\textbf {\bibinfo {volume} {05}},\ \bibinfo
  {pages} {071} (\bibinfo {year} {2011})},\ \Eprint
  {http://arxiv.org/abs/1011.1911} {arXiv:1011.1911 [hep-th]} \BibitemShut
  {NoStop}%
\bibitem [{\citenamefont {Gaiotto}\ \emph {et~al.}(2009)\citenamefont
  {Gaiotto}, \citenamefont {Moore},\ and\ \citenamefont
  {Neitzke}}]{Gaiotto:2009hg}%
  \BibitemOpen
  \bibfield  {author} {\bibinfo {author} {\bibfnamefont {D.}~\bibnamefont
  {Gaiotto}}, \bibinfo {author} {\bibfnamefont {G.~W.}\ \bibnamefont {Moore}},
  \ and\ \bibinfo {author} {\bibfnamefont {A.}~\bibnamefont {Neitzke}},\
  }\href@noop {} {\  (\bibinfo {year} {2009})},\ \Eprint
  {http://arxiv.org/abs/0907.3987} {arXiv:0907.3987 [hep-th]} \BibitemShut
  {NoStop}%
\bibitem [{\citenamefont {Cecotti}\ \emph {et~al.}(2010)\citenamefont
  {Cecotti}, \citenamefont {Neitzke},\ and\ \citenamefont
  {Vafa}}]{Cecotti:2010fi}%
  \BibitemOpen
  \bibfield  {author} {\bibinfo {author} {\bibfnamefont {S.}~\bibnamefont
  {Cecotti}}, \bibinfo {author} {\bibfnamefont {A.}~\bibnamefont {Neitzke}}, \
  and\ \bibinfo {author} {\bibfnamefont {C.}~\bibnamefont {Vafa}},\ }\href@noop
  {} {\  (\bibinfo {year} {2010})},\ \Eprint {http://arxiv.org/abs/1006.3435}
  {arXiv:1006.3435 [hep-th]} \BibitemShut {NoStop}%
\bibitem [{\citenamefont {Cecotti}\ and\ \citenamefont
  {Del~Zotto}(2013)}]{Cecotti:2012jx}%
  \BibitemOpen
  \bibfield  {author} {\bibinfo {author} {\bibfnamefont {S.}~\bibnamefont
  {Cecotti}}\ and\ \bibinfo {author} {\bibfnamefont {M.}~\bibnamefont
  {Del~Zotto}},\ }\href {\doibase 10.1007/JHEP01(2013)191} {\bibfield
  {journal} {\bibinfo  {journal} {JHEP}\ }\textbf {\bibinfo {volume} {01}},\
  \bibinfo {pages} {191} (\bibinfo {year} {2013})},\ \Eprint
  {http://arxiv.org/abs/1210.2886} {arXiv:1210.2886 [hep-th]} \BibitemShut
  {NoStop}%
\bibitem [{\citenamefont {Xie}\ and\ \citenamefont {Zhao}(2013)}]{Xie:2013jc}%
  \BibitemOpen
  \bibfield  {author} {\bibinfo {author} {\bibfnamefont {D.}~\bibnamefont
  {Xie}}\ and\ \bibinfo {author} {\bibfnamefont {P.}~\bibnamefont {Zhao}},\
  }\href {\doibase 10.1007/JHEP03(2013)006} {\bibfield  {journal} {\bibinfo
  {journal} {JHEP}\ }\textbf {\bibinfo {volume} {03}},\ \bibinfo {pages} {006}
  (\bibinfo {year} {2013})},\ \Eprint {http://arxiv.org/abs/1301.0210}
  {arXiv:1301.0210} \BibitemShut {NoStop}%
\bibitem [{\citenamefont {Xie}\ and\ \citenamefont {Yau}(pear)}]{dxie:201601}%
  \BibitemOpen
  \bibfield  {author} {\bibinfo {author} {\bibfnamefont {D.}~\bibnamefont
  {Xie}}\ and\ \bibinfo {author} {\bibfnamefont {S.-T.}\ \bibnamefont {Yau}},\
  }\href@noop {} {\  (\bibinfo {year} {To appear})}\BibitemShut {NoStop}%
\bibitem [{\citenamefont {Boalch}(2008)}]{boalch2008irregular}%
  \BibitemOpen
  \bibfield  {author} {\bibinfo {author} {\bibfnamefont {P.}~\bibnamefont
  {Boalch}},\ }\href@noop {} {\  (\bibinfo {year} {2008})},\ \Eprint
  {http://arxiv.org/abs/0806.1050} {arXiv:0806.1050 [math]} \BibitemShut
  {NoStop}%
\end{thebibliography}%


\end{document}